\documentclass{article}
\usepackage{amssymb}
\usepackage{amsmath}

\begin{document}

\title{Evolution of the $N$-ion Jaynes-Cummings model
beyond the standard rotating wave approximation}
\author{P. Aniello$^{1}$, A. Porzio$^{2}$ \footnote{Alberto Porzio, tel:+39-081-676233;
fax:+39-081-676346, e--mail: alberto.porzio@na.infn.it}
 and S. Solimeno$^{2}$\\
Dipartimento di Scienze Fisiche, Universit\`{a} di Napoli ``Federico II''  \\
and \\
$^{1}$Istituto Nazionale di Fisica Nucleare, Sezione di Napoli \\
$^{2}$''Coherentia'' - INFM, Unit\`{a} di Napoli\\
Complesso Universitario Monte Sant'Angelo,\\
via Cintia, I--80126 Napoli, Italy.}
\maketitle

\begin{abstract}
A unitary transformation of the $N$-ion Jaynes-Cummings Hamiltonian is
proposed. It is shown that any approximate expression of the evolution
operator associated with the \textit{transformed Hamiltonian} retains its
validity independently from the intensity of the external driving field. In
particular, using the rotating wave approximation, one obtains a solution
for the $N$-ion Jaynes-Cummings model which improves the standard rotating
wave approximation solution.
\end{abstract}

\section{Introduction}

The classical Jaynes-Cummings model (JCM)\cite{Jaynes}, with Hamiltonian 
\begin{equation*}
H_{\mathrm{JC}} = \nu\, a^\dagger a + \frac{1}{2}\omega_{ge}\, \sigma_z
+\Omega_R \left(a\, \sigma_+ + a^\dagger\, \sigma_-\right),
\end{equation*}
has become, so to say, the drosophila of quantum optics, since it gives a
simple description of the main properties of many physical systems
consisting of atoms in interaction with electromagnetic fields~\cite{Shore}.
For instance, it has been for long at center of the attention for the study
of laser cooling of ions placed in parabolic traps~\cite{Wineland}\cite
{Bollinger}. The main features of some approximate solutions of the JCM have
been extensively studied, in particular the `collapse and revival'
phenomenon~\cite{Eberly}\cite{Fleischhauer}.

Recently, the interest for the JCM has received a novel impulse in view of
its applications in the fastly developing research area of quantum
computing. Indeed, in a quantum computer (QC), information is stored in a
`quantum register' composed of $N$ two-level systems representing the
quantum bits, or qubits~\cite{Braunstein}. The storage of data and all the
basic operations are implemented by inducing controlled dynamics on the
quantum register~\cite{Kilin}. Since a QC is a quantum mechanical system, it
can perform superpositions of computation operations with a remarkable gain
of efficiency with respect to a classical computer. For instance, it can be
shown that the problem of factoring large numbers into primes, which takes
exponentially increasing time on a classical computer, can be solved in a
polynomially increasing time on a QC~\cite{Shor}.\newline
Now, a couple of electronic states of an ion can be regarded as a concrete
realization of the concept of qubit. Starting from this simple and effective
idea, in 1995 Cirac and Zoller~\cite{Cirac} proposed a concrete model for a
ion-trap computer consisting of $N$ atomic ions trapped in a parabolic
potential well. In this model, each ion is regarded as a two-level system
oscillating between a ground and an excited state ($|g\rangle $ and $%
|e\rangle $), so that the number of available qubits equals the number of
trapped ions. The control of the quantum degrees of freedom is achieved by
addressing the ions with time, frequency and intensity controlled laser
beams (see also~\cite{Pellizzari}). One can consider two possible
arrangements of the laser beams: each ion individually interacting with a
different tightly focalized, possibly zero, laser field (the original Cirac--Zoller
scheme) and the whole set of ions interacting with a single laser
pulse~\cite{Sorensen}. The mutual Coulomb interactions allows a
communication among the trapped ions. The theoretical description of this
physical system is given by a Hamiltonian of the Jaynes-Cummings type ---
which we will call simply the `$N$-ion Jaynes-Cummings Hamiltonian' ---
namely by a Hamiltonian which can be reduced, via a suitable approximation,
to the Hamiltonian of the ($N$-dimensional) JCM. This kind of approximation
is known as the \textit{rotating wave approximation} (RWA)~\cite{Schleich}.
The standard RWA is essentially a perturbative method since its validity
depends on the smallness of a parameter, the Rabi frequency, which is
proportional to the intensity of the laser field. This is a severe limit
since an intense laser field implies a fast coupling between the two
internal energy levels of the trapped ions --- i.e.\ a fast QC --- and any
operation of a ion-trap QC is performed by an accurate control of the
dynamics of the trapped ions.

Hence, to find a solution of the dynamical problem associated with the $N$%
-ion Jaynes-Cummings Hamiltonian, both simple and valid for a wide range of
the intensity of the laser field, is of great importance for the
applications. In the present paper, we have tried to achieve this result.
The basic idea is that of performing \textit{first} suitable transformations
of the Hamiltonian and only \textit{afterwards} applying the RWA (or any
other kind of approximation). Using this strategy, we obtain that the
transformed Hamiltonian is formally similar to the initial one, but the new
perturbative parameter is not proportional any more to the intensity of the
laser field but a simple bounded function of it. Thus the RWA, or any other
approximation, once applied to the transformed Hamiltonian, will be scarcely
sensitive of the laser intensity. In this sense, we have gone beyond the 
\textit{standard} RWA.

The paper is organized as follows. In section~\ref{basic}, the Hamiltonian
describing the system of trapped ions in interaction with a single laser
beam, expressed in a proper set of collective coordinates, is introduced and
the standard RWA is resumed. Then, in section~\ref{beyond}, suitable
transformations are performed on this Hamiltonian in order to obtain a
`balanced Hamiltonian' with respect to the laser intensity. At this point,
in correspondence to the resonances of the system, one is able to find a
good solution to the dynamical problem and this is shown in section~\ref
{evolution}. In section~\ref{extension}, the results are extended to the
general case when each ion interacts with a different laser beam.
Eventually, a discussion of the main results obtained is given in section~%
\ref{discussion}.

\section{The basic ion trap Hamiltonian}
\label{basic}

A system of $N$ equal ions of mass $\mu $ in a potential trap, with strong
confinement along the $y$ and $z$ axes, and weak harmonic binding of
frequency $\nu _{1}$ along the $x$-axis (the `trap axis'), can be described
--- neglecting the motion of the ions transverse to the trap axis --- by a
Hamiltonian of the following type (using units in which $\hbar =1$): 
\begin{equation*}
\sum_{p=1}^{N}\nu _{p}\,a_{p}^{\dagger }a_{p}+W\left( \left\{
a_{p}+a_{p}^{\dagger }\right\} \right) ,
\end{equation*}
where $\{a_{p}\}$ are the annihilation operators associated with a suitable
system of collective coordinates, namely the \textit{normal coordinates} of
the ion chain, and $W\left( \left\{ a_{p}^{{}}+a_{p}^{\dagger }\right\}
\right) $ is the anharmonic component of the ion-ion Coulomb interaction.
Precisely, there is a nonsingular real symmetric matrix $\left[ M_{jp}\right]
$ (see, for instance, ref.~\cite{James}) such that, denoted by $x_{j}$ the
displacement of the $j$-th ion from its equilibrium position, the following
formula holds: 
\begin{equation}
x_{j}=\sum_{p=1}^{N}M_{jp}(a_{p}+a_{p}^{\dagger }).
\end{equation}
Hence, the quantity $x_{jp}\equiv M_{jp}\left( a_{p}+a_{p}^{\dagger }\right) 
$ can be regarded as the displacement of the $j$-th ion due to the
excitation of the $p$-th mode. The first mode of the ion chain, the \textit{%
center of mass mode} associated with the operator $a_{1}$, corresponds
classically to the ion chain oscillating back and forth rigidly. For $N\geq 2
$, the second mode, called the \textit{breathing mode} (or stretch mode),
corresponds classically to the situation in which the center of mass is
motionless while each ion oscillates with an amplitude proportional to its
equilibrium distance from the trap center.

Let us now fix some $j\in \{1,\ldots ,N\}$ and let us suppose that the $j$%
-th ion in the array is addressed by a laser beam of frequency $\omega _{L}$
in a traveling wave configuration. Then, if the $j$-th ion is regarded as a
two-level system oscillating between a ground $|g\rangle $ and an excited
state $|e\rangle $, the ion array will be described by Hamiltonian of the
Jaynes-Cummings type, namely 
\begin{equation}
H(t)=H_{0}+W\left( \left\{ a_{p}+a_{p}^{\dagger }\right\} \right)
+H_{\updownarrow }(t),
\end{equation}
where 
\begin{equation}
H_{0}:=\sum_{p=1}^{N}\nu _{p}\,\hat{n}_{p}+\frac{1}{2}\omega _{ge}\,\sigma
_{z}^{j},
\end{equation}
\begin{equation}
H_{\updownarrow }(t):=\Omega _{R}\left( e^{i\omega _{L}t}\,\sigma _{-}^{j}\,%
\mathcal{D}^{\dagger \,2}+e^{-i\omega _{L}t}\,\sigma _{+}^{j}\,\mathcal{D}%
^{2}\right) ,
\end{equation}
with $\hat{n}_{p}=a_{p}^{\dagger }\,a_{p}^{{}}$ the number operator and $%
\Omega _{R}=\wp \,\mathcal{E}$ the Rabi frequency, $\mathcal{E}$ denoting
the intensity of the laser field. Moreover, we have set: 
\begin{equation}
\mathcal{D}:=\exp \left( \frac{i}{2}\sum_{p=1}^{N}\eta _{jp}\left(
a_{p}+a_{p}^{\dagger }\right) \right) ,
\end{equation}
where 
\begin{equation}
\eta _{jp}:=\frac{k_{L}\cos \phi }{\sqrt{2\mu \nu _{1}}}\,M_{jp}
\end{equation}
--- with $\mathbf{k}_{L}$ the wavevector and $\phi $ the angle between the $x
$-axis and $\mathbf{k_{L}}$ --- plays the role of a Lamb-Dicke factor.
Notice that if $D_{p}(\alpha )$, $\alpha \in \mathbb{C}$, is a displacement
operator associated with the $p$-th mode, namely 
\begin{equation}
D_{p}(\alpha )=\exp \left( \alpha a_{p}^{\dagger }-\alpha^{\ast }
\,a_{p}^{{}}\right) ,\ \ \ D_{p}(\alpha )\,a_{p}\,D_{p}(\alpha )^{\dagger
}=a_{p}-\alpha ,
\end{equation}
we have: 
\begin{equation}
\mathcal{D}=\prod_{p=1}^{N}\exp \left( i\frac{\eta _{jp}}{2}(a_{p}^{\dagger
}+a_{p}^{{}})\right) =\prod_{p=1}^{N}D_{p}\left( i\frac{\eta _{jp}}{2}%
\right) .
\end{equation}
The Hilbert space of the total system (`pointlike' ions + internal degrees
of freedom of the $j$-th ion), namely $\mathcal{H}\otimes \mathbb{C}^{2},\ 
\mathcal{H}\equiv L^{2}(\mathbb{R}^{N})$, is isomorphic to --- and will be
identified with --- the space $\mathcal{H}\oplus \mathcal{H}$, since it is
often useful to work with operator matrices.

Now, the dynamical problem associated with the time-dependent Hamiltonian $%
H(t)$ can be reduced to the dynamical problem for a time-independent
Hamiltonian. Indeed, switching to the interaction picture with reference
Hamiltonian $\frac{1}{2}\omega _{L}t\,\sigma _{z}^{j}$ and setting 
\begin{equation}
R_{t}:=\exp \left( i\frac{1}{2}\omega _{L}t\,\sigma _{z}^{j}\right) ,
\end{equation}
one obtains the time-independent Hamiltonian 
\begin{eqnarray}
\widetilde{H}\!\! &=&\!\!R_{t}\left( H(t)-\frac{1}{2}\omega _{L}\,\sigma
_{z}^{j}\right) R_{t}^{\dagger }  \notag \\
&=&\!\!\sum_{p=1}^{N}\nu _{p}\,\hat{n}_{p}+\frac{1}{2}\delta \,\sigma
_{z}^{j}+\Omega _{R}\left( \sigma _{-}^{j}\,\mathcal{D}^{\dagger \,2}+\sigma
_{+}^{j}\,\mathcal{D}^{2}\right) +\widehat{W},  \label{Hamiltonian}
\end{eqnarray}
where $\delta :=\omega _{ge}-\omega _{L}$ is the ion-laser detuning and we
have set $\widehat{W}\equiv W\left( \left\{ a_{p}+a_{p}^{\dagger }\right\}
\right) $.\newline
At this point, assuming that the anharmonic term $\widehat{W}$ can be
neglected, the standard \textit{rotating wave approximation} consists in
expanding the exponentials that appear in $\mathcal{D}^{2}$, 
\begin{eqnarray*}
\mathcal{D}^{2}\!\! &=&\!\!\prod_{p=1}^{N}\exp \left( i\eta _{jp}\left(
a_{p}^{\dagger }+a_{p}\right) \right)  \\
\!\! &=&\!\!\prod_{p=1}^{N}\left( 1+i\eta _{jp}(a_{p}^{\dagger }+a_{p})-%
\frac{\eta _{jp}^{2}}{2}(a_{p}^{\dagger }+a_{p})^{2}+\ldots \right) ,
\end{eqnarray*}
then passing to the interaction picture with reference Hamiltonian 
\begin{equation*}
\sum_{p=1}^{N}\nu _{p}\,\hat{n}_{p}+\frac{1}{2}\delta \,\sigma _{z}^{j},
\end{equation*}
so obtaining the interaction picture Hamiltonian 
\begin{equation*}
\widetilde{H}_{\mathrm{int}}(t)=\left( 1-i\sum_{p=1}^{N}\eta _{jp}(e^{i\nu
_{p}t}\,a_{p}^{\dagger }+e^{-i\nu _{p}t}\,a_{p})+\ldots \right) e^{-i\delta
t}\,\sigma _{-}^{j}+h.c.\ ,
\end{equation*}
and, finally, retaining only that terms in $\widetilde{H}_{\mathrm{int}}(t)$
which are slowly rotating and at most linear in the Lamb-Dicke parameters,
that in applications are usually assumed to be small ($\eta _{jp}\ll 1$).
Hence, in correspondence to the three types of resonance 
\begin{equation*}
\omega _{ge}-\omega _{L}\simeq 0,\ \ \ \omega _{ge}-\omega _{L}+\nu
_{k}\simeq 0,\ \ \ \omega _{ge}-\omega _{L}-\nu _{k}\simeq 0,
\end{equation*}
one obtains respectively the following three types of effective interaction
picture Hamiltonian: 
\begin{equation*}
\Omega _{R}\left( \sigma _{-}^{j}+\sigma _{+}^{j}\right) ,\ \ \ i\eta
_{jk}\,\Omega _{R}\left( a_{k}^{\dagger }\,\sigma _{+}^{j}-a_{k}\,\sigma
_{-}^{j}\right) ,\ \ \ i\eta _{jk}\,\Omega _{R}\left( a_{k}\,\sigma
_{+}^{j}-a_{k}^{\dagger }\,\sigma _{-}^{j}\right) .
\end{equation*}
The first one, does not couple the internal degrees of freedom of the ions
with the vibrational modes, hence it describes a situation which is not
useful for the quantum computing applications. The last one is precisely the
Jaynes-Cummings Hamiltonian.

We remark that the standard RWA is not always a good approximation. Its
validity can be checked by means of a perturbative expansion of the
evolution operator, for instance by the Magnus expansion~{\cite{Dattoli}}.
In particular, it turns out that it does not work in the strong field regime
(large Rabi frequency $\Omega_R$), since the Lamb-Dicke factor $\eta_{jk}$
does not depend on the \textit{intensity} of the laser field but only on its
frequency, so that the interaction picture Hamiltonian becomes `too large'
for a perturbative treatment.

\section{Beyond the standard rotating wave approximation}
\label{beyond}

In order to go beyond the standard rotating wave approximation, we will show
that is convenient to apply it \textit{after} that the Hamiltonian $%
\widetilde{H}$ has undergone suitable transformations. Aim of these
transformations is to obtain a `good Hamiltonian' which is the sum of a
large component having a simple `diagonal' form --- i.e. such that its
matrix representation in the standard basis 
\begin{equation}  \label{standard}
\!\!\!\{ |n_1\rangle\otimes\cdots\otimes|n_N\rangle\otimes|g\rangle,\
|n_1\rangle\otimes\cdots\otimes|n_N\rangle\otimes|e\rangle :
n_1,\ldots,n_N=1,2,\ldots\}
\end{equation}
is diagonal --- and a small spin-flip component scarcely sensitive of the
Rabi frequency $\Omega_R$.

As a first step, extending the transformation used by Moya-Cessa \textit{et
al.}~\cite{Moya} for a single ion to the N-ion system, namely the
transformation associated with the unitary operator 
\begin{eqnarray}
T_{1}\!\!:= &&\!\!\frac{1}{\sqrt{2}}\left( \frac{1}{2}\left( \mathcal{D}+%
\mathcal{D}^{\dagger }\right) -\frac{1}{2}\left( \mathcal{D}-\mathcal{D}%
^{\dagger }\right) \sigma _{z}^{j}+\mathcal{D}\,\sigma _{+}^{j}-\mathcal{D}%
^{\dagger }\,\sigma _{-}^{j}\right)   \notag \\
\!\! &=&\!\!\frac{1}{\sqrt{2}}\left[ 
\begin{array}{rl}
\mathcal{D}^{\dagger } & \mathcal{D} \\ 
-\mathcal{D}^{\dagger } & \mathcal{D}
\end{array}
\right] ,
\end{eqnarray}
we are able to `linearize' the exponentials that appear in the spin-flip
component of $\widetilde{H}$ (see Eq.(\ref{Hamiltonian})). Indeed, using the
relations 
\begin{equation}
T_{1}\,\hat{n}_{p}\,T_{1}^{\dagger }=\left[ 
\begin{array}{cc}
\hat{n}_{p}+\frac{\eta _{jp}^{2}}{4} & i\frac{\eta _{jp}}{2}%
(a_{p}-a_{p}^{\dagger }) \\ 
i\frac{\eta _{jp}}{2}(a_{p}-a_{p}^{\dagger }) & \hat{n}_{p}+\frac{\eta
_{jp}^{2}}{4}
\end{array}
\right] ,
\end{equation}
\begin{equation}
T_{1}\,\sigma _{z}^{j}\,T_{1}^{\dagger }=-\sigma _{x}^{j},
\end{equation}
\begin{equation}
T_{1}\,\left( \sigma _{-}^{j}\,\mathcal{D}^{\dagger \,2}+\sigma _{+}^{j}\,%
\mathcal{D}^{2}\right) T_{1}^{\dagger }=\sigma _{z}^{j},
\end{equation}
and the fact that $T_{1}$ commutes with $\widehat{W}$, one obtains the
transformed Hamiltonian 
\begin{equation}
\mathfrak{H}=T_{1}\left( \widetilde{H}-\frac{1}{4}\sum_{p=1}^{N}\eta
_{jp}^{2}\,\nu _{p}\right) T_{1}^{\dagger }=\mathfrak{H}_{0}+\mathfrak{W},
\end{equation}
where 
\begin{eqnarray}
\mathfrak{H}_{0}\!\! &=&\!\!\sum_{p=1}^{N}\nu _{p}\,\hat{n}_{p}+\Omega
_{R}\,\sigma _{z}^{j}-\frac{1}{2}\delta \left( \sigma _{-}^{j}+\sigma
_{+}^{j}\right) , \\
\mathfrak{W}\!\! &=&\!\!\widehat{W}+i\sum_{p=1}^{N}\frac{\eta _{jp}\,\nu _{p}%
}{2}\left( a_{p}-a_{p}^{\dagger }\right) \left( \sigma _{-}^{j}+\sigma
_{+}^{j}\right) ,
\end{eqnarray}
and we have dropped the unessential constant term $\frac{1}{4}%
\sum_{p=1}^{N}\eta _{jp}^{2}\,\nu _{p}.$\newline
Let us notice that the spin-flip component of $\mathfrak{W}$ does not depend
on the intensity of the laser field but only on its frequency (through the
Lamb-Dicke parameters). On one hand, this fact is positive in the strong
field regime. On the other hand, in the weak field regime such component
cannot be regarded as a small perturbation. In particular, in the limit for
the field intensity going to zero (hence: $\Omega _{R}\rightarrow 0$), one
observes that $\mathfrak{H}$ does not assume a trivial form, as it happens
for the Hamiltonians $H$ and $\widetilde{H}$. With this problem in mind, let
us perform two further transformations. First, the component $\mathfrak{H}%
_{0}$ can be easily `diagonalized'. Indeed, let us denote by $T_{2}$ a spin
rotation by the angle $\theta $ round the $y$-axis; namely, let us set 
\begin{equation}
T_{2}:=\left[ 
\begin{array}{lr}
\cos \theta /2 & -\sin \theta /2 \\ 
\sin \theta /2 & \cos \theta /2
\end{array}
\right] .
\end{equation}
Now, since 
\begin{eqnarray}
T_{2}\left( \sigma _{-}^{j}+\sigma _{+}^{j}\right) T_{2}^{\dagger }\!\!
&=&\!\!-\sin \theta \,\sigma _{z}^{j}+\cos \theta \left( \sigma
_{-}^{j}+\sigma _{+}^{j}\right) , \\
T_{2}\,\sigma _{z}^{j}\,T_{2}^{\dagger }\!\! &=&\!\!\cos \theta \,\sigma
_{z}^{j}+\sin \theta \,\left( \sigma _{-}^{j}+\sigma _{+}^{j}\right) ,
\end{eqnarray}
we have that, if the angle $-\pi /2\leq \theta \leq \pi /2$ verifies the
condition 
\begin{equation}
\tan \theta =\frac{\Delta }{2},\ \ \ \Delta :=\frac{\omega _{ge}-\omega _{L}%
}{\Omega _{R}}=\frac{\delta }{\Omega _{R}},
\end{equation}
then the following transformations hold: 
\begin{eqnarray}
T_{2}\,\mathfrak{H}_{0}\,T_{2}^{\dagger }\!\! &=&\!\!\sum_{p=1}^{N}\nu _{p}\,%
\hat{n}_{p}+\frac{1}{2}\Omega _{R}\sqrt{4+\Delta ^{2}}\,\sigma _{z}^{j}, \\
T_{2}\,\mathfrak{W}\,T_{2}^{\dagger }\!\! &=&\!\!\widehat{W}-i\frac{\Delta }{%
2\sqrt{4+\Delta ^{2}}}\sum_{p=1}^{N}\eta _{jp}\nu _{p}\left(
a_{p}-a_{p}^{\dagger }\right) \sigma _{z}^{j}  \notag \\
&+&\!\!i\frac{1}{\sqrt{4+\Delta ^{2}}}\sum_{p=1}^{N}\eta _{jp}\nu _{p}\left(
a_{p}-a_{p}^{\dagger }\right) \left( \sigma _{-}^{j}+\sigma _{+}^{j}\right) ,
\end{eqnarray}
Eventually, reorganizing the various terms, we have: 
\begin{equation}
\widehat{H}:=T_{2}\,\mathfrak{H}\,T_{2}^{\dagger }=\widehat{H}_{0}+\widehat{H%
}_{\updownarrow }+\widehat{W},
\end{equation}
where 
\begin{eqnarray}
\widehat{H}_{0}\!\!:= &&\!\!\sum_{p=1}^{N}\nu _{p}\,\hat{n}_{p}+\left( \frac{%
\Omega _{R}}{2}\sqrt{4+\Delta ^{2}}-\frac{i\Delta }{2\sqrt{4+\Delta ^{2}}}%
\sum_{p=1}^{N}\eta _{jp}\,\nu _{p}\left( a_{p}-a_{p}^{\dagger }\right)
\right) \sigma _{z}^{j},\ \ \ \ \  \\
\widehat{H}_{\updownarrow }\!\!:= &&\!\!\frac{i}{\sqrt{4+\Delta ^{2}}}%
\sum_{p=1}^{N}\left( \eta _{jp}\,\nu _{p}\left( a_{p}-a_{p}^{\dagger
}\right) \left( \sigma _{-}^{j}+\sigma _{+}^{j}\right) \right) .
\end{eqnarray}
At this point, in the limit for the laser intensity going to zero, the small
spin-flip component $\widehat{H}_{\updownarrow }$ of $\widehat{H}$ cancels,
as it should. The large component $\widehat{H}_{0}$ still need to be
simplified, since it contains a `nondiagonal' term (with respect to the
standard basis).\newline
To this aim, let us consider the unitary operator 
\begin{eqnarray}
T_{3}\!\!:= &&\!\!\frac{1}{2}\left( \mathcal{D}(\{\alpha _{p}\})+\mathcal{D}%
(\{\alpha _{p}\})^{\dagger }\right) +\frac{1}{2}\left( \mathcal{D}(\{\alpha
_{p}\})-\mathcal{D}(\{\alpha _{p}\})^{\dagger }\right) \sigma _{z}^{j} 
\notag \\
\!\! &=&\!\!\left[ 
\begin{array}{cc}
\mathcal{D}(\{\alpha _{p}\}) & 0 \\ 
0 & \mathcal{D}(\{\alpha _{p}\})^{\dagger }
\end{array}
\right] ,
\end{eqnarray}
with $\mathcal{D}(\{\alpha _{p}\}):=\prod_{p=1}^{N}D_{p}(\alpha _{p})$,
where $\{\alpha _{p}\}$ is a set of c-numbers to be fixed in a suitable way.
Then, observe that 
\begin{eqnarray*}
T_{3}^{{}}\,\widehat{H}_{0}\,T_{3}^{\dagger }\!\! &=&\!\!\sum_{p=1}^{N}\nu
_{p}\left( \hat{n}_{p}-\left( \alpha _{p}^{\ast }\,a_{p}+\alpha
_{p}\,a_{p}^{\dagger }\right) \sigma _{z}^{j}+|\alpha _{p}|^{2}\right) +%
\frac{1}{2}\Omega _{R}\sqrt{4+\Delta ^{2}}\,\sigma _{z}^{j} \\
\!\! &-&\!\!i\frac{\Delta }{2\sqrt{4+\Delta ^{2}}}\sum_{p=1}^{N}\eta
_{jp}\,\nu _{p}\left( \left( a_{p}-a_{p}^{\dagger }\right) \sigma
_{z}^{j}-\left( \alpha _{p}-\alpha _{p}^{\ast }\right) \right) .
\end{eqnarray*}
Thus, if we assume that 
\begin{equation}
\alpha _{p}=i\frac{\Delta }{2\sqrt{4+\Delta ^{2}}}\,\eta _{jp},
\end{equation}
we get: 
\begin{equation}
T_{3}\,\widehat{H}_{0}\,T_{3}^{\dagger }=\sum_{p=1}^{N}\nu _{p}\,\hat{n}_{p}+%
\frac{1}{2}\Omega _{R}\sqrt{4+\Delta ^{2}}\,\sigma _{z}^{j}-\frac{\Delta ^{2}%
}{4(4+\Delta ^{2})}\sum_{p=1}^{N}\eta _{jp}\,\nu _{p},
\end{equation}
namely the `diagonalized' expression that we wished to obtain. Moreover, we
have: 
\begin{eqnarray}
T_{3}\,\left( a_{p}-a_{p}^{\dagger }\right) \left( \sigma _{-}^{j}+\sigma
_{+}^{j}\right) \,T_{3}^{\dagger }\!\! &=&\!\!\left( a_{p}-a_{p}^{\dagger
}\right) \left( \sigma _{-}^{j}\,\breve{\mathcal{D}}^{\dagger \,2}+\sigma
_{+}^{j}\,\breve{\mathcal{D}}^{2}\right)   \notag \\
\!\! &+&\!\!i\frac{\Delta }{\sqrt{4+\Delta ^{2}}}\,\eta _{jp}\left( \sigma _{-}^{j}\,\breve{\mathcal{D}}^{\dagger \,2}-\sigma
_{+}^{j}\,\breve{\mathcal{D}}^{2}\right) ,
\end{eqnarray}
where 
\begin{equation}
\breve{\mathcal{D}}:=\prod_{p=1}^{N}D_{p}\left( i\frac{\breve{\eta}_{jp}}{2}%
\right) ,\ \ \ \breve{\eta}_{jp}:=\frac{\Delta }{\sqrt{4+\Delta ^{2}}}\,\eta
_{jp}.
\end{equation}
Notice that, since 
\begin{eqnarray}
\lim_{\Omega _{R}\rightarrow 0}\breve{\eta}_{jp}\!\! &=&\!\!\mathrm{sign}%
(\delta )\,\eta _{jp},\ \ \ \ \ \ \ \ (\mbox{weak field limit}) \\
\lim_{\Omega _{R}\rightarrow \infty }\breve{\eta}_{jp}\!\! &=&\!\!0,\ \ \ \
\ \ \ \ \ \ \ \ \ \ \ \ \ \ \ (\mbox{strong field limit})
\end{eqnarray}
$\breve{\eta}_{jp}$ behaves as a `laser intensity balanced' Lamb-Dicke
parameter.\newline
Hence, we have the transformed Hamiltonian 
\begin{equation}
\breve{H}:=T_{3}\left( \widehat{H}+\frac{\Delta ^{2}}{4(4+\Delta ^{2})}%
\sum_{p=1}^{N}\eta _{jp}^{2}\,\nu _{p}\right) T_{3}^{\dagger }=\breve{H}_{0}+%
\breve{H}_{\updownarrow }+\widehat{W},
\end{equation}
where again we have dropped the unessential constant term and we have set 
\begin{equation}
\breve{H}_{0}:=\sum_{p=1}^{N}\nu _{p}\,\hat{n}_{p}+\frac{1}{2}\breve{\delta}%
\,\sigma _{z}^{j},
\end{equation}
with 
\begin{equation}
\breve{\delta}:=\sqrt{4\Omega _{R}^{2}+\delta ^{2}},
\end{equation}
and 
\begin{eqnarray}
\breve{H}_{\updownarrow }\!\!:= &&\!\!\frac{i}{\Delta }\sum_{p=1}^{N}\breve{%
\eta}_{jp}\,\nu _{p}\left( a_{p}-a_{p}^{\dagger }\right) \left( \sigma
_{-}^{j}\,\breve{\mathcal{D}}^{\dagger \,2}+\sigma _{+}^{j}\,\breve{\mathcal{%
D}}^{2}\right)   \notag \\
\!\! &-&\!\!\frac{1}{\Delta }\sum_{p=1}^{N}\breve{\eta}_{jp}^{2}\,\nu
_{p}\left( \sigma _{-}^{j}\,\breve{\mathcal{D}}^{\dagger \,2}-\sigma
_{+}^{j}\,\breve{\mathcal{D}}^{2}\right) .
\end{eqnarray}
The new Hamiltonian $\breve{H}$ enjoys very nice properties. The first point
is that the main component $\breve{H}_{0}$ has a very simple diagonal form.
It only differs from the main component of $\widetilde{H}$ for the constant
factor 
\begin{equation*}
\frac{1}{2}\breve{\delta}=\frac{1}{2}\sqrt{4\Omega _{R}^{2}+(\omega
_{ge}-\omega _{L})^{2}}
\end{equation*}
of $\sigma _{z}^{j}$ instead of 
\begin{equation*}
\frac{1}{2}\delta =\frac{1}{2}(\omega _{ge}-\omega _{L}).
\end{equation*}
Thus, we have obtained an effective ion-laser detuning $\breve{\delta}$
which will introduce a correction into the standard condition for resonances
(obviously, the standard formula is recovered in the weak field limit).%
\newline
The second --- but not less important --- point is that the spin-flip
component has a good behaviour in both the weak and strong field regime;
indeed: 
\begin{equation}
\lim_{\Omega _{R}\rightarrow 0}\breve{H}_{\updownarrow }=0
\end{equation}
and, since $\lim_{\Delta \rightarrow 0}\breve{\eta}_{jp}/\Delta =\eta _{jp}/2
$, 
\begin{equation}
\lim_{\Omega _{R}\rightarrow \infty }\breve{H}_{\updownarrow }=\frac{i}{2}%
\sum_{p=1}^{N}\eta _{jp}\,\nu _{p}\left( a_{p}-a_{p}^{\dagger }\right)
\left( \sigma _{-}^{j}+\sigma _{+}^{j}\right) .
\end{equation}
Moreover, if the Lamb-Dicke parameters $\{\eta _{jp}\}$ are assumed to be
small, $\eta _{jp}\ll 1$, since 
\begin{equation*}
|\breve{\eta}_{jp}|\leq |\eta _{jp}|\ \ \Rightarrow \ \ \breve{\eta}_{jp}\,%
\breve{\mathcal{D}}^{2}\simeq \breve{\eta}_{jp},
\end{equation*}
\begin{equation*}
\frac{1}{\sqrt{4+\Delta ^{2}}}>\frac{\Delta }{4+\Delta ^{2}}\ \ \Rightarrow
\ \ \frac{1}{\Delta }\,\breve{\eta}_{jp}\gg \frac{1}{\Delta }\,\breve{\eta}%
_{jp}^{2},
\end{equation*}
one has that, \textit{for any intensity of the laser field}, 
\begin{equation}
\breve{H}_{\updownarrow }\simeq \frac{i}{\Delta }\sum_{p=1}^{N}\breve{\eta}%
_{jp}\,\nu _{p}\left( a_{p}-a_{p}^{\dagger }\right) \left( \sigma
_{-}^{j}+\sigma _{+}^{j}\right) .
\end{equation}

\section{The evolution operator}

\label{evolution}

In this section, we will exploit the transformations introduced above in
order to give a suitable approximate expression of the evolution operator $%
U(H;t,t_0)$ associated with the ion trap Hamiltonian $H$: 
\begin{equation}
i \frac{d}{dt}\, U(H; t,t_0)= H(t)\, U(H; t,t_0),\ \ \ U(H;t_0,t_0)= \mathrm{%
Id}.
\end{equation}
Parenthetically, let us notice that, if 
\begin{equation}
\bar{H}(t)=H_0 +\widehat{W}+\Theta(t)\,H_\updownarrow(t),
\end{equation}
where $\Theta$ denotes the Heaviside theta function, i.e.\ if $\bar{H}$ is
the Hamiltonian describing the situation in which the interaction with the
laser field is turned on at $t=0$, then, for $t_0<0<t$, we have simply: 
\begin{equation}
U(\bar{H};t,t_0)= U(H;t,0)\, \exp\left(i(H_0+\widehat{W})t_0\right).
\end{equation}

First, we will assume that the ions are not too tightly packed along the
trap axis so that we can make the approximation 
\begin{equation}
\breve{H}\simeq \breve{H}_{0}+\breve{H}_{\updownarrow },
\label{approximation}
\end{equation}
namely we can neglect the anharmonic component $\widehat{W}$ of the ion-ion
potential. This assumption, whose accuracy can be estimated calculating the
equilibrum positions of the ions in the trap (see~\cite{James}), is also
necessary in order to prevent the ion chain from adopting a zig-zag
configuration~\cite{Schiffer}. Next, we can pass to the interaction picture
with reference Hamiltonian $\breve{H}_{0}$. Then, if we set 
\begin{equation}
V_{t}:=\exp \left( i\breve{H}_{0}\,t\right) ,
\end{equation}
the effective Hamiltonian of the system becomes: 
\begin{eqnarray}
J\!\!C(t)\!\!:= &&\!\!V_{t}\,\breve{H}_{\updownarrow }\,V_{t}^{\dagger } 
\notag \\
\!\! &=&\!\!\frac{i}{\Delta }\sum_{p=1}^{N}\breve{\eta}_{jp}\,\nu _{p}\left(
e^{-i\omega _{p}^{-}\,t}\,a_{p}\,\sigma _{+}^{j}\,\breve{\mathcal{D}}%
_{t}-e^{i\omega _{p}^{-}\,t}\,a_{p}^{\dagger }\,\sigma _{-}^{j}\,\breve{%
\mathcal{D}}_{t}^{\dagger }\right)   \notag \\
\!\! &+&\!\!\frac{i}{\Delta }\sum_{p=1}^{N}\breve{\eta}_{jp}\,\nu _{p}\left(
e^{-i\omega _{p}^{+}\,t}\,a_{p}\,\sigma _{-}^{j}\,\breve{\mathcal{D}}%
_{t}^{\dagger }-e^{i\omega _{p}^{+}\,t}\,a_{p}^{\dagger }\,\sigma _{+}^{j}\,%
\breve{\mathcal{D}}_{t}\right)   \notag \\
\!\! &-&\!\!\frac{1}{\Delta }\sum_{p=1}^{N}\breve{\eta}_{jp}^{2}\,\nu
_{p}\left( e^{-i\breve{\delta}\,t}\,\sigma _{-}^{j}\,\breve{\mathcal{D}}%
_{t}^{\dagger }-e^{i\breve{\delta}\,t}\,\sigma _{+}^{j}\,\breve{\mathcal{D}}%
_{t}\right) ,  \label{JC1}
\end{eqnarray}
where we have set 
\begin{equation}
\omega _{p}^{\pm }:=\nu _{p}\pm \breve{\delta}=\nu _{p}\pm \sqrt{4\Omega
_{R}^{2}+\left( \omega _{ge}-\omega _{L}\right) ^{2}},
\end{equation}
and 
\begin{equation}
\breve{\mathcal{D}}_{t}:=\prod_{p=1}^{N}\exp \left( i\breve{\eta}_{jp}\left(
e^{-i\nu _{p}\,t}\,a_{p}+e^{i\nu _{p}\,t}\,a_{p}^{\dagger }\right) \right) .
\end{equation}
In correspondence to the resonance $\omega _{k}^{-}\simeq 0$, if $\eta
_{jk}\ll 1$, applying the RWA, one finds that 
\begin{equation}
J\!\!C(t)\simeq \frac{i}{\Delta }\,\breve{\eta}_{jk}\,\nu _{k}\left(
a_{k}\,\sigma _{+}^{j}-a_{k}^{\dagger }\,\sigma _{-}^{j}\right)   \label{app}
\end{equation}
and hence: 
\begin{equation}
U(J\!\!C;t,t_{0})\simeq \exp \left( \frac{\breve{\eta}_{jk}\,\nu _{k}}{%
\Delta }(t-t_{0})\left( a_{k}\,\sigma _{+}^{j}-a_{k}^{\dagger }\,\sigma
_{-}^{j}\right) \right) .
\end{equation}
Then, since 
\begin{equation*}
\left( a_{k}\,\sigma _{+}^{j}-a_{k}^{\dagger }\,\sigma _{-}^{j}\right)
^{2m}=(-1)^{m}\left( a_{k}\,a_{k}^{\dagger }\,\sigma _{+}^{j}\,\sigma
_{-}^{j}+a_{k}^{\dagger }\,a_{k}\,\sigma _{-}^{j}\,\sigma _{+}^{j}\right)
^{m},
\end{equation*}
it follows that 
\begin{eqnarray*}
U(J\!\!C;t,t_{0})\!\! &\simeq &\!\!\sum_{m=0}^{\infty }\frac{(-1)^{m}}{2m!}%
\left( \frac{\breve{\eta}_{jk}\,\nu _{k}}{\Delta }(t-t_{0})\right) ^{2m}%
\left[ \!\!
\begin{array}{cc}
(a_{k}\,a_{k}^{\dagger })^{m} & 0 \\ 
0 & (a_{k}^{\dagger }\,a_{k})^{m}
\end{array}
\!\!\right]  \\
\!\! &+&\!\!\sum_{m=0}^{\infty }\frac{(-1)^{m}}{(2m+1)!}\left( \frac{\breve{%
\eta}_{jk}\,\nu _{k}}{\Delta }(t-t_{0})\right) ^{2m+1}\left[ \!\!
\begin{array}{cc}
0 & (a_{k}\,a_{k}^{\dagger} )^{m}a_{k} \\ 
-(a_{k}^{\dagger }\,a_{k})^{m}a_{k}^{\dagger } & 0
\end{array}
\!\!\right] .
\end{eqnarray*}
Thus, within the approximation~{(\ref{app})}, the evolution operator $%
U(J\!\!C;t,t_{0})$ has the following form: 
\begin{equation}
\left[ 
\begin{array}{cc}
\cos \left( \frac{\breve{\eta}_{jk}\,\nu _{k}}{\Delta }(t-t_{0})\sqrt{\hat{n}%
_{k}+1}\right)  & \dfrac{\sin \left( \frac{\breve{\eta}_{jk}\,\nu _{k}}{%
\Delta }(t-t_{0})\sqrt{\hat{n}_{k}+1}\right) }{\sqrt{\hat{n}_{k}+1}}\,a_{k}
\\ 
-\dfrac{\sin \left( \frac{\breve{\eta}_{jk}\,\nu _{k}}{\Delta }(t-t_{0})%
\sqrt{\hat{n}_{k}}\right) }{\sqrt{\hat{n}_{k}}}\,a_{k}^{\dagger } & \cos
\left( \frac{\breve{\eta}_{jk}\,\nu _{k}}{\Delta }(t-t_{0})\sqrt{\hat{n}_{k}}%
\right) 
\end{array}
\right] .  \label{matrix}
\end{equation}

At this point, we will use the the transformations we performed on the
original Hamiltonian $H$ in order to obtain the expression of the evolution
operator $U(H;t,t_{0})$. Let us resume them: 
\begin{equation*}
H(t)\ \ \overset{R_{t}}{\dashrightarrow }\ \ \widetilde{H}\ \ \overset{T_{1}%
}{\longrightarrow }\ \ \mathfrak{H}\ \ \overset{T_{2}}{\longrightarrow }\ \ 
\widehat{H}\ \ \overset{T_{3}}{\longrightarrow }\ \ \breve{H}\simeq \breve{H}%
_{0}+\breve{H}_{\updownarrow }\ \ \overset{V_{t}}{\dashrightarrow }\ \
J\!\!C(t),
\end{equation*}
where the dashed arrows denote the switching to an interaction picture. Now,
taking care of the difference between genuine unitary transformations of the
Hamiltonian and Dirac (interaction picture) decompositions of the evolution
operator, we have: 
\begin{eqnarray*}
U(H;t,t_{0})\!\! &=&\!\!R_{t}^{\dagger }\,\exp \left( -i\left(
t-t_{0}\right) \widetilde{H}\right)  \\
\!\! &=&\!\!R_{t}^{\dagger }\,T_{1}^{\dagger }\,\exp \left( -i\left(
t-t_{0}\right) \mathfrak{H}\right) \,T_{1} \\
\!\! &=&\!\!R_{t}^{\dagger }\,T_{1}^{\dagger }\,T_{2}^{\dagger }\,\exp
\left( -i\left( t-t_{0}\right) \widehat{H}\right) \,T_{2}\,T_{1} \\
\!\! &=&\!\!R_{t}^{\dagger }\,T_{1}^{\dagger }\,T_{2}^{\dagger
}\,T_{3}^{\dagger }\,\exp \left( -i\left( t-t_{0}\right) \breve{H}\right)
\,T_{3}\,T_{2}\,T_{1}.
\end{eqnarray*}
Then, using the approximation $\breve{H}\simeq \breve{H}_{0}+\breve{H}%
_{\updownarrow }$, we get: 
\begin{eqnarray}
U(H;t,t_{0})\!\! &\simeq &\!\!R_{t}^{\dagger }\,T_{\Delta }^{\dagger }\,\exp
\left( -i\left( t-t_{0}\right) \breve{H}_{0}+\breve{H}_{\updownarrow
}\right) \,T_{\Delta }  \notag \\
\!\! &=&\!\!R_{t}^{\dagger }\,T_{\Delta }^{\dagger }\,V_{t}^{\dagger
}\,U(J\!\!C;t,t_{0})\,T_{\Delta },
\end{eqnarray}
where we have set 
\begin{eqnarray}
\!\!\!T_{\Delta }\!\!:= &&\!\!T_{3}\,T_{2}\,T_{1}  \notag \\
\!\! &=&\!\!\!\!\!\left[ \!\!
\begin{array}{cl}
\varkappa _{\Delta }^{+}\,\prod_{p=1}^{N}D_{p}(i\epsilon _{\Delta
}^{-}\,\eta _{jp}) & \varkappa _{\Delta
}^{-}\,\prod_{p=1}^{N}D_{p}(i\epsilon _{\Delta }^{+}\,\eta _{jp}) \\ 
-\varkappa _{\Delta }^{-}\,\prod_{p=1}^{N}D_{p}(i\epsilon _{\Delta
}^{+}\,\eta _{jp})^{\dagger } & \varkappa _{\Delta
}^{+}\,\prod_{p=1}^{N}D_{p}(i\epsilon _{\Delta }^{-}\,\eta _{jp})^{\dagger }
\end{array}
\!\!\right]   \notag \\
\!\! &=&\!\!\!\!\!\left[ \!\!
\begin{array}{cl}
\varkappa _{\Delta }^{+}\,\prod_{p=1}^{N}D_{p}\left( i(\breve{\eta}%
_{jp}-\eta _{jp})/2\right)  & \varkappa _{\Delta
}^{-}\,\prod_{p=1}^{N}D_{p}\left( i(\breve{\eta}_{jp}+\eta _{jp})/2\right) 
\\ 
-\varkappa _{\Delta }^{-}\,\prod_{p=1}^{N}D_{p}\left( i(\breve{\eta}%
_{jp}+\eta _{jp})/2\right) ^{\dagger } & \varkappa _{\Delta
}^{+}\,\prod_{p=1}^{N}D_{p}\left( i(\breve{\eta}_{jp}-\eta _{jp})/2\right)
^{\dagger }
\end{array}
\!\!\right] ,
\end{eqnarray}
with 
\begin{equation}
\varkappa _{\Delta }^{\pm }=\sqrt{\frac{1}{4}+\frac{1}{2\sqrt{4+\Delta ^{2}}}%
}\,\pm \,\mathrm{sign}(\Delta )\,\sqrt{\frac{1}{4}-\frac{1}{2\sqrt{4+\Delta
^{2}}}}\ ,
\end{equation}
\begin{equation}
\epsilon _{\Delta }^{\pm }=\frac{\Delta }{2\sqrt{4+\Delta ^{2}}}\,\pm \,%
\frac{1}{2}\ .
\end{equation}

\section{Extension: N laser beams}

\label{extension}

The results obtained above can be easily generalized to the case when each
ion in the trap interacts with a laser beam at the same time. Notice that
the case when all the ions are interacting simultaneously with the same
laser beam will be included as a particular case.

If each ion interacts with a laser beam, then the internal degrees of
freedom of all the ions will be involved in the dynamics of the system.
Hence, the Hilbert space of the model will be 
\begin{equation*}
\mathcal{H}\otimes \mathbb{C}^{2N}\cong \overbrace{\left( \mathcal{H}\oplus 
\mathcal{H}\right) \otimes \ldots \otimes \left( \mathcal{H}\oplus \mathcal{H%
}\right) }^{N},
\end{equation*}
where $\mathcal{H}\equiv L^{2}(\mathbb{R}^{N})$, while the Hamiltonian will
be now the following: 
\begin{eqnarray}
H(t)\!\! &=&\!\!\sum_{p=1}^{N}\nu _{p}\,\hat{n}_{p}+\sum_{j=1}^{N}\frac{1}{2}%
\omega _{ge}\,\sigma _{z}^{j}  \notag \\
\!\! &+&\!\!\sum_{j=1}^{N}\Omega _{R}^{j}\left( e^{-i(\omega
_{L}^{j}t+\varphi ^{j})}\,\sigma _{+}^{j}\,\mathcal{D}_{j}^{2}+e^{i(\omega
_{L}^{j}t+\varphi ^{j})}\,\sigma _{-}^{j}\,\mathcal{D}_{j}^{\dagger
\,2}\right) +\widehat{W},
\end{eqnarray}
where $\Omega _{R}^{j}$, $\omega _{L}^{j}$ are respectively the Rabi and the
laser frequency associated with the $j$-th laser beam (namely the one in
interaction with the $j$-th ion), $\varphi ^{j}$ is an initial phase, and we
have set 
\begin{equation}
\mathcal{D}_{j}^{2}:=\prod_{p=1}^{N}\exp \left( i\eta _{jp}\left(
a_{p}^{\dagger }+a_{p}\right) \right) =\prod_{p=1}^{N}D_{p}^{2}\left( i\frac{%
\eta _{jp}}{2}\right) ,
\end{equation}
$\left[ \eta _{jp}\right] =\left[ \left( k_{L}^{j}\cos \phi _{j}/\sqrt{2\mu
\nu _{1}}\right) M_{jp}\right] $ being a `Lamb-Dicke matrix'. Let us perform
on the Hamiltonian $H(t)$ the following transformations: 
\begin{equation*}
H(t)\ \ \overset{R_{t}}{\dashrightarrow }\ \ \widetilde{H}\ \ %
\xrightarrow{T_{\{\Delta_j\}}^{}}\ \ \breve{H}=\breve{H}_{0}+\breve{H}%
_{\updownarrow }+\widehat{W}\simeq \breve{H}_{0}+\breve{H}_{\updownarrow }\
\ \overset{V_{t}}{\dashrightarrow }\ \ J\!\!C(t),
\end{equation*}
where we recall that the dashed arrows denote a passage to an interaction
picture and we have: 
\begin{equation}
R_{t}:=\bigotimes_{j=1}^{N}\exp \left( \frac{i}{2}\left( \omega
_{L}^{j}t+\varphi ^{j}\right) \sigma _{z}^{j}\right) ,
\end{equation}
\begin{equation}
T_{\{\Delta _{j}\}}:=\bigotimes_{j=1}^{N}\left[ 
\begin{array}{cl}
\varkappa _{\Delta _{j}}^{+}\,\prod_{p=1}^{N}D_{p}(i\epsilon _{\Delta
_{j}}^{-}\,\eta _{jp}) & \varkappa _{\Delta
_{j}}^{-}\,\prod_{p=1}^{N}D_{p}(i\epsilon _{\Delta _{j}}^{+}\,\eta _{jp}) \\ 
-\varkappa _{\Delta _{j}}^{-}\,\prod_{p=1}^{N}D_{p}(i\epsilon _{\Delta
_{j}}^{+}\,\eta _{jp})^{\dagger } & \varkappa _{\Delta
_{j}}^{+}\,\prod_{p=1}^{N}D_{p}(i\epsilon _{\Delta _{j}}^{-}\,\eta
_{jp})^{\dagger }
\end{array}
\right] ,
\end{equation}
with 
\begin{equation}
\Delta _{j}:=\frac{\omega _{ge}-\omega _{L}^{j}}{\Omega _{R}^{j}},
\end{equation}
\begin{equation}
\varkappa _{\Delta _{j}}^{\pm }:=\sqrt{\frac{1}{4}+\frac{1}{2\sqrt{4+\Delta
_{j}^{2}}}}\,\pm \,\mathrm{sign}(\Delta _{j})\,\sqrt{\frac{1}{4}-\frac{1}{2%
\sqrt{4+\Delta _{j}^{2}}}}\ ,
\end{equation}
\begin{equation}
\epsilon _{\Delta _{j}}^{\pm }:=\frac{\Delta _{j}}{2\sqrt{4+\Delta _{j}^{2}}}%
\,\pm \,\frac{1}{2}\ .
\end{equation}
Accordingly, the expression of the transformed Hamiltonian $\breve{H}=\breve{%
H}_{0}+\breve{H}_{\updownarrow }$ is given by 
\begin{equation}
\breve{H}_{0}=\sum_{p=1}^{N}\nu _{p}\,\hat{n}_{p}+\frac{1}{2}\sum_{j=1}^{N}%
\breve{\delta}_{j}\,\sigma _{z}^{j},
\end{equation}
\begin{eqnarray}
\breve{H}_{\updownarrow }\!\! &=&\!\!\sum_{j,p=1}^{N}\frac{i}{\Delta _{j}}\,%
\breve{\eta}_{jp}\,\nu _{p}\left( a_{p}-a_{p}^{\dagger }\right) \left(
\sigma _{-}^{j}\,\breve{\mathcal{D}}_{j}^{\dagger \,2}+\sigma _{+}^{j}\,%
\breve{\mathcal{D}}_{j}^{2}\right)   \notag \\
\!\! &-&\!\!\sum_{j,p=1}^{N}\frac{1}{\Delta _{j}}\,\breve{\eta}%
_{jp}^{2}\,\nu _{p}\left( \sigma _{-}^{j}\,\breve{\mathcal{D}}_{j}^{\dagger
\,2}-\sigma _{+}^{j}\,\breve{\mathcal{D}}_{j}^{2}\right) ,
\end{eqnarray}
with 
\begin{equation}
\breve{\delta}_{j}:=\sqrt{4\Omega _{R}^{j\,2}-\left( \omega _{ge}-\omega
_{L}^{j}\right) ^{2}},\ \ \ \breve{\eta}_{jp}:=\frac{\Delta _{j}}{\sqrt{%
4+\Delta _{j}^{2}}}\,\eta _{jp},
\end{equation}
\begin{equation}
\breve{\mathcal{D}}_{j}:=\prod_{p=1}^{N}D_{p}\left( i\frac{\breve{\eta}_{jp}%
}{2}\right) .
\end{equation}
Then, we have 
\begin{equation}
V_{t}:=\exp \left( i\breve{H}_{0}\,t\right) 
\end{equation}
and the expression of interaction picture Hamiltonian $J\!\!C$ in the
general case is the following: 
\begin{eqnarray}
J\!\!C(t)\!\! &=&\!\!V_{t}\,\breve{H}_{\updownarrow }\,V_{t}^{\dagger } 
\notag \\
\!\! &=&\!\!\sum_{j,p=1}^{N}\frac{i}{\Delta _{j}}\,\breve{\eta}_{jp}\,\nu
_{p}\left( e^{-i\omega _{j,p}^{-}\,t}\,a_{p}\,\sigma _{+}^{j}\,\breve{%
\mathcal{D}}_{j;t}-e^{i\omega _{j,p}^{-}\,t}\,a_{p}^{\dagger }\,\sigma
_{-}^{j}\,\breve{\mathcal{D}}_{j;t}^{\dagger }\right)   \notag \\
\!\! &+&\!\!\sum_{j,p=1}^{N}\frac{i}{\Delta _{j}}\,\breve{\eta}_{jp}\,\nu
_{p}\left( e^{-i\omega _{j,p}^{+}\,t}\,a_{p}\,\sigma _{-}^{j}\,\breve{%
\mathcal{D}}_{j;t}^{\dagger }-e^{i\omega _{j,p}^{+}\,t}\,a_{p}^{\dagger
}\,\sigma _{+}^{j}\,\breve{\mathcal{D}}_{j;t}\right)   \notag \\
\!\! &-&\!\!\sum_{j,p=1}^{N}\frac{1}{\Delta _{j}}\,\breve{\eta}%
_{jp}^{2}\,\nu _{p}\left( e^{-i\breve{\delta}_{j}\,t}\,\sigma _{-}^{j}\,%
\breve{\mathcal{D}}_{j;t}^{\dagger }-e^{i\breve{\delta}_{j}\,t}\,\sigma
_{+}^{j}\,\breve{\mathcal{D}}_{j;t}\right) ,  \label{JC2}
\end{eqnarray}
where we have set 
\begin{equation}
\omega _{j,p}^{\pm }:=\nu _{p}\pm \breve{\delta}_{j}=\nu _{p}\pm \sqrt{%
4\Omega _{R}^{j\,2}+\left( \omega _{ge}-\omega _{L}^{j}\right) ^{2}}
\end{equation}
and 
\begin{equation}
\breve{\mathcal{D}}_{j;t}:=\prod_{p=1}^{N}\exp \left( i\breve{\eta}%
_{jp}\left( e^{-i\nu _{p}\,t}\,a_{p}+e^{i\nu _{p}\,t}\,a_{p}^{\dagger
}\right) \right) .
\end{equation}
Thus, a resonance arises in the interaction picture Hamiltonian $J\!\!C$ if
the condition $\omega _{j,p}^{-}=0$ is satisfied for some $j,p\in \{1,\ldots
,N\}$. Notice that up to $N$ resonances may arise simultaneously, with up to 
$N$ vibrational modes being involved. The evolution operator, in the RWA, is
given by a tensor product of identity operators and single mode evolution
operators like~{(\ref{matrix})} (one for each resonant mode).

\section{Discussion}

\label{discussion}

A very frequent problem in quantum mechanics is to find an approximate
expression of the evolution operator associated with a Hamiltonian of the
following type: 
\begin{equation*}
H(t)=H_{0}+H_{\updownarrow }(t;\{\varepsilon _{k}\}),
\end{equation*}
where $H_{0}$ is a explicitly diagonalizable, in the sense that it admits a
complete set of known eigenvectors, and $\{\varepsilon _{k}\}$ is a set of
adjustable parameters. Typically, $H_{\updownarrow }(t;\{\varepsilon
_{k}\})=0$, for $t<0$, and the initial ($t=0$) state vector of the system is
an eigenvector of $H_{0}$ (namely, for $t<0$, a stationary state of $H$).
Then, a perturbative treatment of the problem is possible only if $%
H_{\updownarrow }(t;\{\varepsilon _{k}\})$ is `comparatively small' with
respect to $H_{0}$. Usually, such condition holds only for a restricted
range of the adjustable parameters $\{\varepsilon _{k}\}$. This turns out to
be a severe limit in many remarkable applications.\newline
A problem of this sort affects the standard RWA treatment of a Hamiltonian
of the Jaynes-Cummings type describing a set of linearly trapped ions in
interaction with laser beams, if the laser intensity is not small. In the
present paper, we have shown that such problem can be bypassed. This is
accomplished by performing suitable transformations of the Hamiltonian.
These transformations allow us to obtain an interaction picture Hamiltonian
which is \textit{stable} with respect to the intensity of the laser field.
Indeed, recalling formula~{(\ref{JC1})} (or~{(\ref{JC2})}), one observes
that the dependence on the Rabi frequency is contained in the factors 
\begin{equation*}
\frac{1}{\Delta }\,\breve{\eta}_{jp}=\frac{\Omega _{R}}{\sqrt{4\Omega
_{R}^{2}+\delta ^{2}}}\,\eta _{jp},
\end{equation*}
with $0\leq \left| \frac{1}{\Delta }\,\breve{\eta}_{jp}\right| \leq \frac{1}{%
2}\,|\eta _{jp}|$, and 
\begin{equation*}
\frac{1}{\Delta }\,\breve{\eta}_{jp}^{2}=\frac{\Omega _{R}\,\delta }{4\Omega
_{R}^{2}+\delta ^{2}}\,\eta _{jp}^{2},
\end{equation*}
with $0\leq \left| \frac{1}{\Delta }\right| \breve{\eta}_{jp}^{2}\leq \frac{1%
}{4}\,\eta _{jp}^{2}$. Hence the validity of any approximate expression of
the evolution operator, in particular the one obtained using the RWA, will
be scarcely influenced by the intensity of the laser field.\newline
There are three further points that should be highlighted. First, a simple
field-intensity corrected condition for resonances arises in a natural way
from our procedure, namely (in the single laser beam case): 
\begin{equation*}
\nu _{p}-\breve{\delta}=0,\ \ \ \breve{\delta}=\sqrt{4\Omega _{R}^{2}+\left(
\omega _{ge}-\omega _{L}\right) ^{2}}.
\end{equation*}
Observe that the corrected detuning $\breve{\delta}$ is always nonnegative,
differently from the standard ion-laser detuning $\delta $. Thus, both
positive-$\delta $ and negative-$\delta $ resonances will be associated with
terms of the type 
\begin{equation*}
a_{p}\,\sigma _{+}^{j},\ \ \ a_{p}^{\dagger }\,\sigma _{-}^{j},
\end{equation*}
in the interaction picture Hamiltonian $J\!\!C$, hence formally treated in
the same way, differently from the standard RWA.\newline
Second, one can check that the action on the standard basis~{(\ref{standard})%
} of the unitary operator $T_{\Delta }$ (or $T_{\{\Delta _{j}\}}$) is easily
computable. This means simple explicit expressions for the time evolution of
state vectors. A detailed analysis of the main features of these expressions
and, in paricular, of the typical `collapse and revival' phenomenon will be
presented in a subsequent paper~{\cite{Aniello1}}. \newline
Third, the nonharmonic component $\widehat{W}$ of the ion-ion interaction is
not modified by any of the transformations performed. To show this we have
retained it up to formula~{(\ref{approximation})}, where it is neglected.
Hence its relatively simple form is not corrupted and it can be taken into
account as a perturbation in a more accurate treatment. Again, such a
treatment will be presented in a subsequent paper~{\cite{Aniello2}}.

Other interesting unitary transformations have been presented, with
different aims, in the literature~\cite{Jonathan} \cite{Arevalo} \cite{You},
apart from the previously cited one~\cite{Moya}. They do not seem to share
the same properties with the one presented here.

\section*{Acknowledgements}

The main results of this paper were presented by one of the authors
(P.~Aniello) during the `Wigner Centennial Conference' held in P\'ecs,
Hungary (8-12 July 2002). He wishes to thank the organizers for the kind
hospitality.

\end{document}